\newif\ifmarkup\markupfalse

\documentclass[english, longbibliography, preprint]{revtex4-2}
\usepackage[T1]{fontenc}
\usepackage[latin9]{inputenc}
\usepackage{color}
\usepackage[english]{babel}
\usepackage{amssymb}
\usepackage{graphicx}
\usepackage{amsmath}
\usepackage{breakurl}
\usepackage{graphicx}
\usepackage{verbatim}
\usepackage[unicode=true,pdfusetitle, bookmarks=true,bookmarksnumbered=false,bookmarksopen=false, breaklinks=false,pdfborder={0 0 1},backref=false,colorlinks=true]{hyperref}
\hypersetup{linkcolor=blue, citecolor=blue, urlcolor  = blue}
\graphicspath{ {./figures/} }

\newcommand{\Reynolds}{{\rm Re}}
\newcommand{\Strouhal}{{\rm St}}


\begin{document}

\title{The effect of menisci on vortex streets on soap film flows}
\author{Ildoo Kim}
\email{ildoo.kim.phys@gmail.com}
\affiliation{Department of Mechatronics, Konkuk University, Chungju, South Korea 27478}
\date{\today}

\begin{abstract}
In soap film experiments, the insertion of an external object is necessary to produce vorticity.
However, this insertion causes local thickness changes, or simply {\it meniscus}, near the object.
Because the meniscus formation may alter the flow near the object, the characterization of meniscus is of considerable importance for the accurate interpretation of data.
In this study, we insert cylindrical cones made of aluminum, titanium, and glass to measure the size of the menisci by using a long range microscope.
In all material tested, we find that the size of meniscus is less than 0.2 mm, much shorter than the capillary length.
In addition, by comparing the formation of vortex streets behind objects of different materials, we conclude that the meniscus acts as an added length to the size of the object itself.
This added length effect can be non-negligible if the size of object is comparable to the size of meniscus.
\end{abstract}

\maketitle
\newpage

\section{Introduction}

The flowing soap film channel \cite{Gharib:1989um,Chomaz:1990vl,Kellay:1995wk} is an attractive tool of experiment in testing the formation, self-organization, and evolution of vortices in \ifmarkup\color{red}\fi two-dimensional flow, \color{black} because it is easy to handle, convenient to visualize, and inexpensive.
Because of these advantages, there are many research papers using soap film channels to investigate various phenomena including the wake structure \cite{Kim:2015jp,Kim:2019kw,10.1103/physrevfluids.4.114802}, fluid-structure interaction \cite{Jung:2006tn,Singh:2015kv}, and turbulence \cite{Martin:1998ty,Rutgers:1998uz,Kellay:2012tl,Cerbus:2013js,Kim:2021en}.

Experiments with soap films typically involve inserting an external object into the soap film.
In many cases, we use soap film channels because we are ultimately interested in the interactions of vortices in two-dimensional \ifmarkup\color{red}\fi flow, \color{black} e.g. turbulence \cite{Kraichnan:1967us,Kraichnan:1980uy,Martin:1998ty}.
Therefore, vorticity is one of the main keywords in the majority of soap film experiments, and the insertion of an object is used as a means to produce vorticity.
For example, a cylindrical cone is inserted into the soap films to investigate the laminar vortex street in two dimensions \cite{Kim:2015jp}, and a comb is inserted to investigate two-dimensional turbulence \cite{Martin:1998ty}.

%

When an external object is inserted, the local thickness of the film can change because of the formation of meniscus.
But, in many studies, it is casually and implicitly assumed that the inserted object do not alter the flow but only creates vorticity.
This is commonly believed so that the experiments with soap films make general sense to the community.
The presence of meniscus may affect the downstream flow in two scenarios. 
In the first scenario, Marangoni stress may occur because of the thickness change at the meniscus.
The soap film is thicker at the meniscus, and the surface tension may be weaker than surroundings depending on the soap concentration and mean thickness \cite{Sane:2018uo}.
In this case, the Marangoni stress may act toward the object and alter the flow pattern.
In the second scenario, the meniscus may change the local viscosity.
According to previous studies \cite{Trapeznikov,Prasad:2009jj}, the film viscosity is a sum of the surface viscosity $\nu_s$ and bulk viscosity $\nu_b$, i.e., $\nu=\nu_b+2\nu_s/\delta$, where $\delta$ is the film thickness. 
Therefore, the measurement of the Reynolds number $\Reynolds=UD/\nu$, where $U$ and $D$ are the characteristic velocity and length scale of the system and $\nu$ is the kinematic viscosity, can be inaccurately estimated.

These two scenarios challenge the assumption that the inserted object does not alter the flow, but the literature provides only partial answers.
In 2010, Tran {\it et al.} measured the velocity near the bounding wires of a soap film channel as close as 50 $\rm\mu{m}$ using Laser Doppler velocimetry but observed no deviation from the constant velocity gradient \cite{Tran:2010hn}.
Their observation is consistent with earlier lower-resolution measurements by Rutgers {\it et al.}; they observed no anomalies in the velocity gradient near the bounding wires \cite{Rutgers:1996vg}.
In static soap films, which is considerably thinner than flowing soap films, the formation of menisci around micron-sized fibers was investigated \cite{Guo.2019} in the context of the contact line relaxation \cite{kkr}.
These studies indirectly suggest that the local thickening of meniscus may not play an important role in quantifying the flow structure in soap film flows.
However, the direct observation of menisci and quantitative examination of their effect in flowing soap film setup are yet to be performed.

In this study, we directly observe the formation of menisci and investigate their effect on downstream flow by using two key experimental features.
First, we use a long-distance working microscope to monitor the small object inserted into soap films.
This setup allows us to precisely measure the size of an object and meniscus around it. 
Second, we use objects composed of different materials; aluminum, titanium, and glass.
These materials have different wetting properties with soapy water, and we expect the size of the meniscus to depend on the type of material used.
We postulate that if the meniscus does not affect the downstream flow, the measurements of the spatial and temporal arrangements of the vortex streets created by different materials are indistinguishable.
In particular, the spatial arrangement is quantified by the wavelength, the distance between two vortices in the longitudinal direction.
The wavelength is known to be linearly proportional to $D$, i.e., $\lambda=\alpha D+ \lambda_0$, where $\alpha$ and $\lambda_0$ are parameters \cite{Roushan:2005un}.
The temporal arrangement is quantified by the Strouhal number, $\Strouhal=fD/U$, where $f$ is the shedding frequency.
If the meniscus casts a non-negligible effect on the flow downstream, it should appear in the measurements of $\lambda$ and $\Strouhal$.

We learn from our experiments that the size of the meniscus depends on the size of object $D$, and it affects $\lambda$ and $\Strouhal$. 
The size of the meniscus is proportional to $D$ when $D$ is small ($D<1$ mm) but is bounded by the upper limit, which is approximately 0.2 mm.
Therefore, the effect of meniscus is prominent only when $D$ is small ($\simeq 0.2$ mm) and negligible otherwise.

\section{Experimental Setup}

\begin{figure}
\begin{centering}
\includegraphics[width=8.5cm]{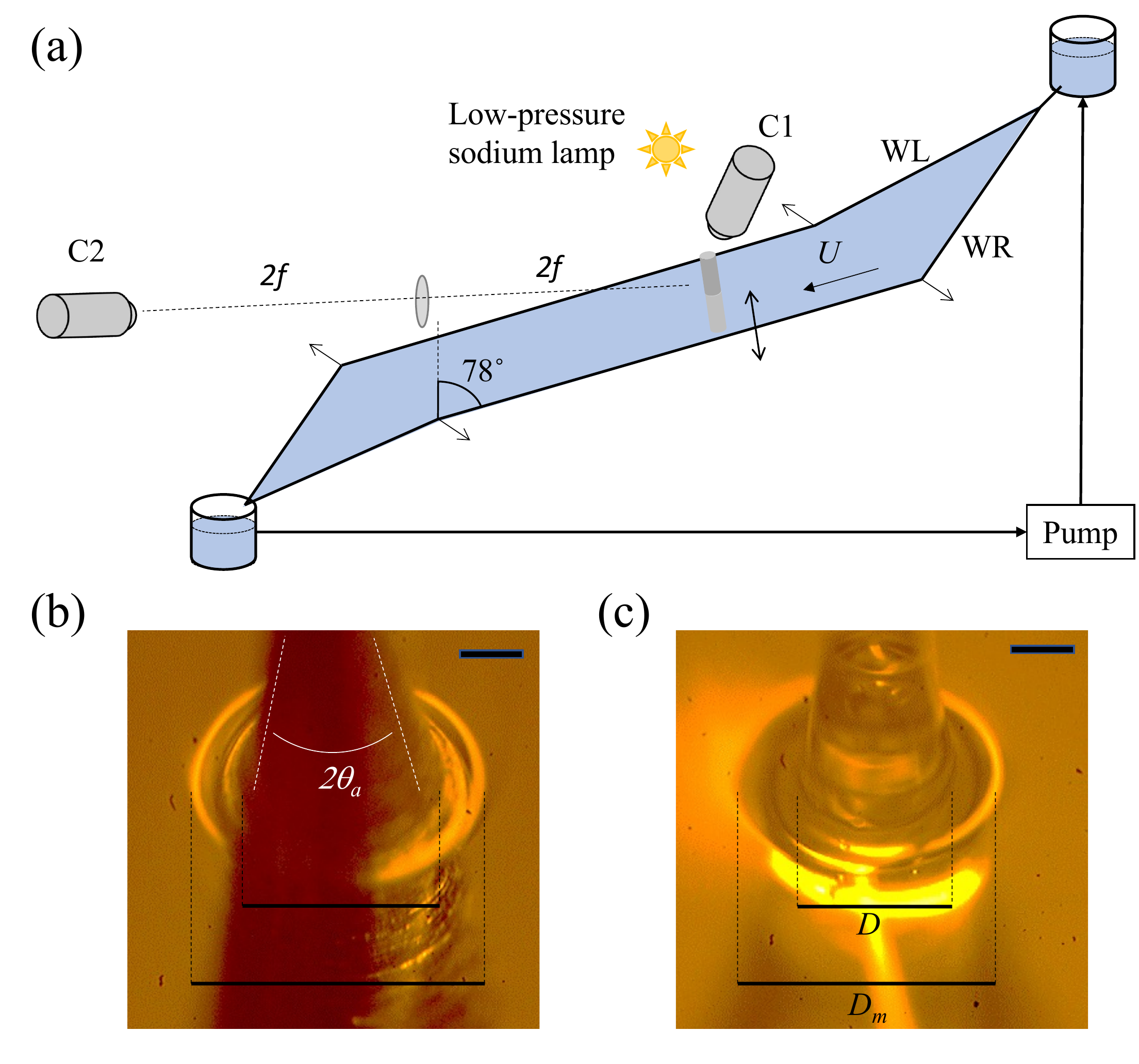}
\par
\end{centering}
\caption{
(a) Experimental setup. 
We use an inclined soap film channel to reduce the flow speed.
The fast video camera (C1) monitors the flow structure on soap film through interferometry, and the long-distance microscope (C2) monitors the insertion depth of the object.
(b) Typical microscopic images of titanium cone and (c) glass cone.
In the image, the size of the object, $D$, and the size including meniscus, $D_m$, are clearly distinguishable, as indicated.
The scale bars in (b) and (c) are 200 $\rm\mu{m}$.
\label{fig:apparatus}}
\end{figure}

Our experiments are conducted using an inclined soap film channel, which is depicted in Fig. \ref{fig:apparatus}(a) and discussed in previous publications \cite{Georgiev:2002kg,Kim:2015jp}.
Briefly, the channel consists of two parallel nylon wires, WL and WR.
These wires are approximately 2 m long and separated by 50 mm, when the flow is established in the soap film channel.
The channel is inclined by 78$^\circ$ to reduce the effect of gravity by 80\%. 
The inclination allows us to access the relatively slow mean flow speed of $U\sim0.65\,\rm m/s$.
At this speed, the thickness of the film is approximately $3 \,\rm \mu m$, yielding a weak compressibility as the Mach-like number is approximately 0.17. \cite{Kim:2017dn}

The soap solution used in the experiments is the 2\% solution of commercial dish soap (Dawn non-ultra, Proctor \& Gamble) to distilled water (the {\it overall} concentration of surfactant $c_0=0.02$).
This solution has a kinematic viscosity of $1.3\times10^{-6} \,\rm m^2/s$ when measured by a capillary tube viscometer.

We insert an external object normal to the soap film, which creates a vortex street downstream.
The external object has a circular cross-section and is tapered such that the Reynolds number, $\Reynolds=UD/\nu$, where $D$ is the diameter of the object, is adjusted by changing the insertion depth.
The test section is approximately 1 m below the top of the channel, and the flow speed does not change much within the range of observation (several centimeters) at this location.

We make three cones made of different materials, as summarized in Table \ref{tab:exp_setup}.
First, we use aluminum cones.
These cones are tapered by a careful machining process up to a tip size of 50 $\rm\mu m$; sharpening the tip further is extremely difficult because of material properties of Al.
Second, we use a titanium cone.
Titanium is also difficult to lathe, because it is harder and more brittle than Al. 
Fortunately, we were able to fabricate one with a tip size of 30 $\rm\mu m$.
These two metallic cones have the apex angle $2\theta_a$ of approximately 26$^\circ$ [see Fig. \ref{fig:apparatus}(a)], and therefore the angle between the film's normal and the cone's sidelines is invariant and equals the half of the apex angle $\theta_a$.
Third, we used a glass cone.
The glass cone is produced using a glass puller, up to a tip size of 10 $\rm \mu m$.
The glass cone's apex angle varies by the insertion depth; $\theta_a$ approaches 10$^\circ$ near the base of the cone, but it decreases to approximately 5$^\circ$ near the tip of the cone.

\begin{table}
\begin{centering}
\begin{tabular}{c|c c c c c}
\hline 
No. & Material & Tip size & $\theta_a$ & $\theta_c$ & $\theta=\theta_a+\theta_c$\\
\hline 
\hline 
1 & Al & 50 $\rm \mu m$ & 13$^\circ$ & 58$\pm$2$^\circ$ & 71$^\circ$\\
\hline 
2 & Ti & 30 $\rm \mu m$ & 13$^\circ$ & 51$\pm$3$^\circ$ & 64$^\circ$\\
\hline 
3 & Glass & 10 $\rm \mu m$ & 5$^\circ$ to 10$^\circ$ & $\approx$0$^\circ$ & $\simeq 5^\circ$\\
\hline 
\end{tabular}
\par\end{centering}
\caption{List of external objects used in the experiments, along with the wetting properties: the apex angle $\theta_a$, the contact angle $\theta_c$, and $\theta\equiv\theta_a+\theta_c$.
The contact angle is measured by the sessile droplet method using 0.05\% soap solution. 
\label{tab:exp_setup}}
\end{table}

While the glass is hydrophilic and having a near-zero contact angle, the metal's wetting properties are measured in-house using the sessile droplet method.
We measure the contact angle between pure water and Al is 87$\pm$1$^\circ$ and that between pure water and Ti is 93$\pm$1$^\circ$.
We remark that these values are somewhat different from the values in literature \cite{10.2109/jcersj2.117.1285}, where they reported 10$^\circ$ to 60$^\circ$ for the contact angle between pure water and Al with oxidation layer and 50$^\circ$ for the contact angle between pure water and titanium with oxidation layer.
The contact angles between soapy water and metals are smaller than that between pure water and metals, as surfactant improves the wettability. 
However, it is difficult to specify the contact angles because they depend on the area-volume ratio of the fluid.
In other words, when a droplet and a soap film are made of the same volume of soapy water, the soap film will have less surface concentration of surfactant thus having a larger surface tension, because of its larger surface area. 
Therefore, 
for the best estimate of contact angles, we need to match the surface concentration rather than the overall concentration of surfactant. 
For the concentration matching, we use the Langmuir adsorption isotherm \cite{Sane:2018uo}
\begin{equation}
\Gamma=\frac{\Gamma^\infty}{1+c^*/c},
\label{eq:Langmuir}
\end{equation}
where $\Gamma$ is the surface concentration and $c$ is the bulk concentration of surfactant.
Combined with the conservation of matter
\begin{equation}
c_0=c+r_{av}\Gamma,
\end{equation}
where $r_{av}$ is the area/volume ratio, to derive the relationship between $\Gamma$ and $c_0$ up to $r_{av}$.
We take the values from the literature: $\Gamma_\infty=0.76\%\cdot \rm\mu{m}$, $c^*=0.1\%$ for films \cite{Sane:2018uo}, and $c^*=0.01\%$ for sessile droplets (in-house measurement), and we conclude that the sessile droplet of $c_0=0.0005$ (0.05\%) has the wetting properties that is closest to that of the soap film made of $c_0=0.02$.
Our measurement using soap solution of $c_0=0.0005$ suggests that the contact angle $\theta_c$ between our soap film and Al and between the soap film and Ti are 
58$\pm$2$^\circ$ and 51$\pm$3$^\circ$, respectively.


We use two cameras to monitor the soap film channel.
The first camera is a high-speed camera [Phantom V5, C1 in Fig. \ref{fig:apparatus}(a)] mounted above the soap film channel to monitor the flow structure downstream the object.
The flow structure is visualized as an interferogram using a monochromatic light source (low-pressure sodium lamp, 589 nm).
The soap film is slightly compressible, and the thickness variation is strongly correlated with the vorticity \cite{Rivera:1998tw}.
This feature allows us to easily capture the flow structure using the interferogram.
The second camera is mounted on a long-distance microscope [C2 in Fig. \ref{fig:apparatus}(a)].
The working distance of a regular microscope is extended using a 2f optical arrangement to approximately 1 m.
This setup allows us to monitor the insertion depth of the object to the soap film and size of object $D$ without hindering the view of the high-speed camera.

Figures \ref{fig:apparatus}(b-c) show typical images from the long-distance microscope, depicting cones intruding the soap films.
The images clearly show a meniscus formed between the object and soap film. 
While the exact profiling of the film thickness is technically challenging, the size of the meniscus is optically defined as the difference between the sizes of the external halo $D_m$ and of the object itself $D$. 
Then, the (lateral) size of meniscus $d$ is calculated as
\begin{equation}
d=\frac{D_m-D}{2}.
\label{eq:meniscus_definition}
\end{equation}

In the latter part, we present the measurements of structural parameters of vortex streets, i.e., the wavelength $\lambda$ and the shedding frequency $f$ of vortex streets in soap films. 
These measurement of $\lambda$ and $f$ are made using video recording from the fast video camera (C1).
The wavelength is defined as a distance between the same-signed vortices, and in fact this distance is a function of downstream distance.
It increases until it reaches an asymptotic value as the vortex street moves downstream \cite{Kim:2015jp}, and we take the asymptotic value as our measurement of $\lambda$.
The frequency $f$ is measured by counting the number of vortices produced in a unit time.

\section{Results and Discussions}

\subsection{Measurement of menisci}

\begin{figure}
\begin{centering}
\includegraphics[viewport=50 20 700 540, width=8cm]{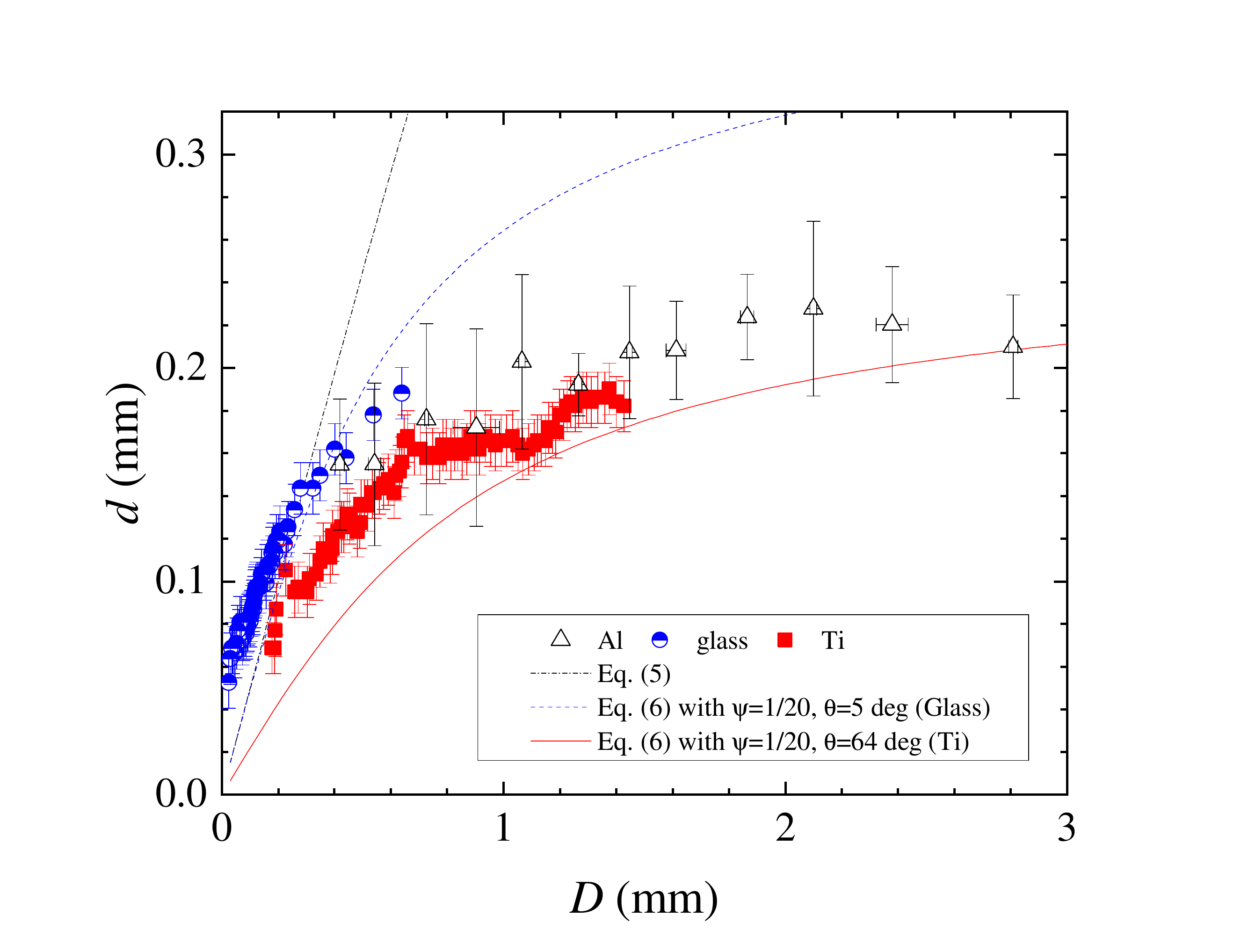}
\par
\end{centering}
\caption{
The measurement of $d$, the size of the meniscus, with respect to $D$.
We collected data using three different cones made of aluminum (open triangles), glass (half-filled circles), and titanium (closed squares). 
The meniscus is formed much smaller than the prediction of the conventional theory in Eq. \eqref{eq:simple_solution}.
The ad-hoc model in Eq. \eqref{eq:adhoc_solution} with the fuzzy parameter \ifmarkup\color{red}\fi $\psi=1/4.5$ is roughly consistent with \color{black} the data.
We note that the margin of error of aluminum cones is larger than the others because of the use of less magnified microscope.
\label{fig:meniscus_measurement}}
\end{figure}

Figure \ref{fig:meniscus_measurement} shows the measurement of $d$ for different materials. 
For all materials, $d$ increases with $D$ when $D$ is relatively small, but it reaches each respective asymptotic values as $D$ gets relatively large.
For aluminum, $d$ increases with $D$ when $D<1\,\rm mm$ but reaches an asymptotic value at 0.22 mm.
For titanium, $d$ increases with $D$ when $D\lesssim1\,\rm mm$ but does not change much when $D>0.07\,\rm mm$, having almost constant value at 0.16 mm.
For glass, our experimental range covers only small $D$; thus the asymptotic value of $d$ is not directly observed, but it shows a clear indication that $d$ approaches an asymptotic value as $D$ increases.

Note that the range of experiments of $D$ are determined by the tip size, which is determined by the material properties in the fabrication process.
For example, aluminum is relatively soft with high ductility and is therefore extremely difficult to cut sharp.
Therefore, the Al cone is the most obtuse, and the measurements using Al range from approximately 0.5 mm to 4 mm.
By constrast, the glass cone is produced by heating and pulling. 
This method allows us to produce extreme sharpness, and we could then inspect $D$ less than 1 mm.

To rationalize our measurement, we first consider a rough and classical dimensional analysis, where the Laplace pressure is balanced with the pressure difference caused by the capillary rise by $h$. 
Formally, we \ifmarkup\color{red}\fi write \color{black}
\begin{equation}
\rho g h \simeq \sigma\left( -\frac{1}{D/2}+\kappa  \cos\theta\right),
\label{eq:basic_eq}
\end{equation}
where $\sigma$ is the surface tension of soap film, \ifmarkup\color{red}\fi $\theta=\theta_a+\theta_c$ is the sum of the apex angle and the contact angle, \color{black} and $\kappa$ is the curvature of the meniscus, and it is conventionally assumed that $\kappa\simeq h^{-1}\simeq d^{-1}$.
\ifmarkup\color{red}\fi
Then, the solution of Eq. \eqref{eq:basic_eq} is as follows:
\begin{equation}
h=l_c\cdot \frac{l_c}{D}\left(\sqrt{1+\cos\theta\frac{D^2}{l_c^2}}-1 \right)\simeq d.
\label{eq:simple_solution}
\end{equation}
\color{black}
Eq. \eqref{eq:simple_solution} separated into two regimes. 
First, when $D$ is relatively small, the effect of gravity is neglected, and the solution solely determined by the balance between two terms in the right-hand side, indicating that the meniscus is proportional to the object size, i.e., $h=D\cos\theta/2$.
The second regime emerges as $D$ increases.
In this regime, the relative importance of $\rho gh$ increases while the relative importance of $D^{-1}$ decreases, and the solution approaches to an asymptotic value $h=l_c\sqrt{\cos\theta}$, where $l_c=\sqrt{\sigma/\rho g}\approx1.7$ mm is the capillary length for the soap films ($\sigma\approx 28$ mN/m \cite{Sane:2018uo}).
However, \ifmarkup\color{red}\fi Eq. \eqref{eq:simple_solution} \color{black} overestimates $d$ when $D$ is large. 
In the large $D$ regime, Eq. \eqref{eq:simple_solution} indicates that $d\rightarrow l_c$, but we find that our measurement of $d$ is much smaller than $l_c$.

While a quantitative theory that fully accommodates our observation is not available at this point, 
we speculate that $d$ is much smaller than $l_c$ because this is the measurement of dynamic meniscus.
The transient profile of meniscus near a thin fiber (diameter of 9 $\rm\mu m$) protruding a static soap film has been investigated by Guo {\it et al.} \cite{Guo.2019}, and they found that the equilibrium shape is reached around $\sim \mathcal{O}(2)$ s.
This time to equilibrium can be shorter if the fiber is thicker, according to the earlier study by
Clanet and Qu\'{e}r\'{e} using Hexane \cite{kkr}.
They find that the time to reach the equilibrium decreases with the increase of the fiber thickness but eventually reaches to a constant value.
In our case, the object size is in the order of $\mathcal{O}(2)$ $\rm\mu{m}$, about ten times larger than the fiber size examined by Guo {\it et al.} \cite{Guo.2019}, therefore we expect a shorter time to equilibrium.
However, the time scale of meniscus formation is only $\frac{D}{U}\simeq1.5$ ms and is much smaller than the findings of Guo {\it et al.}.
Therefore, the measurement in Fig. \ref{fig:meniscus_measurement} should be considered to be dynamic and is out of scope the conventional equilibrium theory.

For ensuing discussion, we introduce an ad-hoc model just to fit the measurement. 
By adding a fuzzy factor $\psi$, we can engineer Eq. \eqref{eq:simple_solution} to cover the whole range of $D$ as follows:
\ifmarkup\color{red}\fi
\begin{equation}
d=\psi l_c \cdot \frac{\psi l_c}{D}\left(\sqrt{1+{\cos\theta}\frac{D^2}{(\psi l_c)^2}}-1 \right).
\label{eq:adhoc_solution}
\end{equation}
\color{black}
Then, we get $d=D\cos\theta/2$ and \ifmarkup\color{red}\fi$d=\psi l_c\sqrt{\cos\theta}$ \color{black} at small and large $D$ regime, respectively.
The transition between two regime occurs near \ifmarkup\color{red}\fi $D\sim2 \psi l_c/\sqrt{\cos\theta}$. \color{black}
We find that \ifmarkup\color{red}\fi $\psi=1/4.5$ is roughly consistent with data, \color{black} as shown in solid curves in Fig. \ref{fig:meniscus_measurement}.

\subsection{Wavelength of vortex streets}

To investigate the effect of the meniscus on vortex formation, we measure the wavelength and inspect its relationship with $D$ and $D_m$. 
In a previous study using glass rods in a vertically flowing soap film channel, Roushan and Wu \cite{Roushan:2005un} has shown that the wavelength $\lambda$ of the vortex street is linearly proportional to $D$ with a nonzero intercept, following the form 
\begin{equation}
\lambda=\lambda_0+ \alpha D ,
\end{equation}
and the measurement of two parameters by Roushan and Wu (RW) are: $\lambda_0^{(RW)}=0.35\pm0.08 \,\rm mm$ and $\alpha^{(RW)}=4.1\pm0.3$.

\begin{figure}
\begin{centering}
\includegraphics[viewport=50 20 700 580, width=8cm]{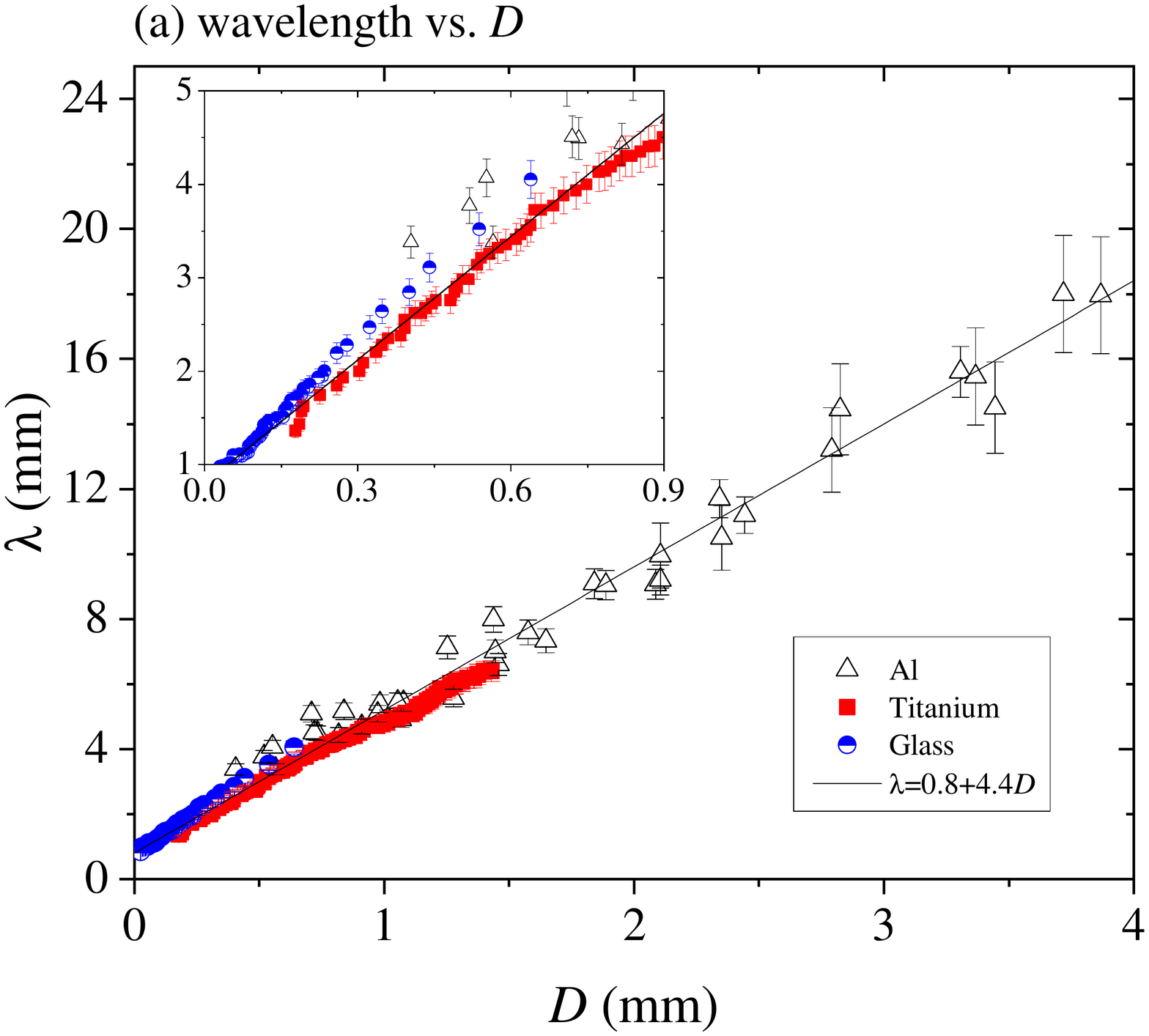}
\includegraphics[viewport=50 20 700 580, width=8cm]{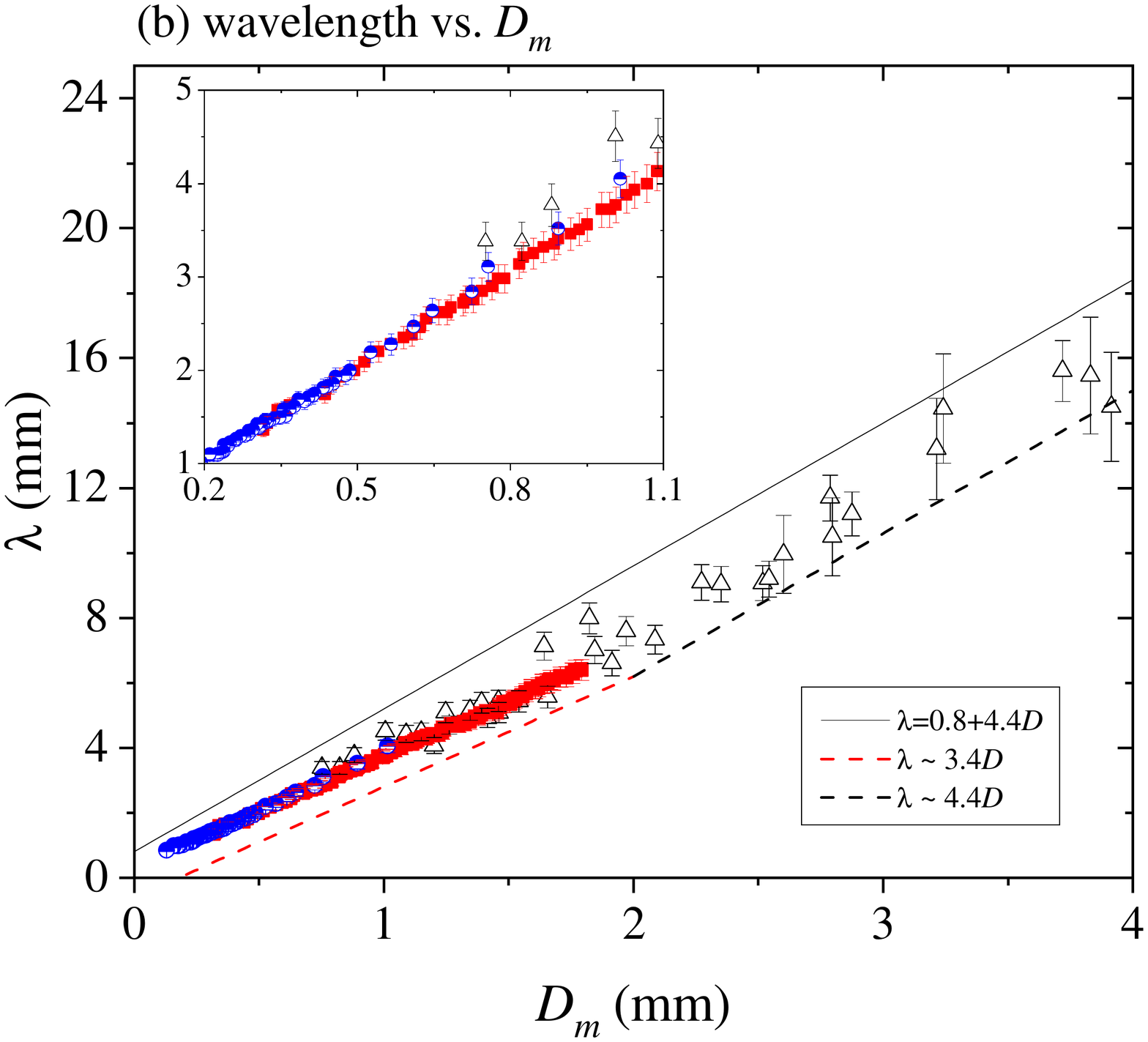}
\par
\end{centering}
\caption{
The measurements of $\lambda$ displayed with respect to (a) $D$ and (b) $D_m$.
While the overall linearity between $\lambda$ and $D$ appears to hold, the data are not continuous, as seen in the inset. 
When plotted with $D_m$, the linearity holds only when $D_m$ is sufficiently large, but the difference between materials is decreased.
\label{fig:wavelength}}
\end{figure}

Figure \ref{fig:wavelength}(a) shows the measurement of the wavelength with respect to $D$, using Al, Ti, and Glass surfaces, showing that the linear relation holds albeit approximately. 
We determine the parameters differently from Roushan and Wu's determination \cite{Roushan:2005un}, as we obtain $\alpha=4.4$ and $\lambda_0=0.8 \,{\rm mm}$.
The proportionality constant matches within the uncertainty margin, but the intercept is approximately two factors larger than their measurement. 
We speculate that the difference in the intercept is caused by the difference in flow speed (RW: 1.2 m/s vs. this work: 0.65 m/s) of the two experiments, because an independent experiment \cite{Kim:2015jp} using the flow speed of 0.6 m/s reports $\lambda_0=1.00\pm0.03 \,\rm mm$ and $\alpha=4.3\pm0.1$.

Although the overall trend of the relationship between $\lambda$ and $D$ is roughly linear, the data from different material are continuous only when $\lambda$ is plotted with respect to $D_m$. 
In the insets of Figs. \ref{fig:wavelength}(a-b), the data are enlarged to show them in detail. 
The inset of Fig. \ref{fig:wavelength}(a) shows that the data acquired using glass and titanium are not connected to each other.
However, in the inset of Fig. \ref{fig:wavelength}(b), the two data are perfectly continuous.
This observation implies that $D_m$ is more relevant length-scale than $D$ in determining $\lambda$.

We remark that the phenomenological linearity of $\lambda$ to $D$ does not apply if data are plotted with respect to $D_m$.
It is now roughly a two-regime curve that is separated into the small $D$ regime and large $D$ regime.
In the small $D$ regime, i.e., \ifmarkup\color{red}\fi $D<2 \psi  l_c/\sqrt{\cos\theta}\simeq1.4$ mm, \color{black} 
we obtain $D={D_m}/{(1+\cos\theta)}$.
From this substitution, the proportionality constant $\alpha$ of $\lambda$-$D$ relationship becomes $\alpha/(1+\cos\theta)$ in $\lambda$-$D_m$ relationship.
Using $\theta\simeq 71^\circ$, the new proportionality constant is approximately 3.3, which matches with our measurement of 3.4.
In the large $D$ regime, 
the proportionality constant remain unchanged, but the intercept does change to \ifmarkup\color{red}\fi $\lambda_0-2\alpha \psi l_c \sqrt{\cos\theta}$. \color{black}
The shift, \ifmarkup\color{red}\fi $2\alpha \psi l_c \sqrt{\cos\theta}$, \color{black} is computed to be 2 mm in the current study and is consistent with the measurement.

In summary, from our experimental observations, we infer the followings.
1) The wavelength of the vortex streets, $\lambda$, is determined by $D_m$, rather than $D$.
2) As the size of the meniscus, $d$, approaches an asymptotic value, the relationship between $\lambda$ and $D$ becomes apparently linear, particularly when the experiments are conducted with relatively large $D$.
3) In two-dimensional soap films, the meniscus should be considered a part of the intruding object.

\subsection{St-Re relationship}

To further confirm our observation of the meniscus affecting vortex formation, we compare the relationships between the Strouhal and Reynolds numbers based on $D$ and $D_m$.
The Strouhal number is defined by $\Strouhal=fD/U$, where $f$ is the vortex shedding frequency, and the Reynolds number is defined by $\Reynolds=UD/\nu$, where $\nu$ is the kinematic viscosity of the fluid.
These definitions are modified using $D_m$ instead of $D$, i.e., $\Strouhal_m=fD_m/U$ and $\Reynolds_m=UD_m/\nu$.

\begin{figure}
\begin{centering}
\includegraphics[viewport=50 20 700 540, width=8cm]{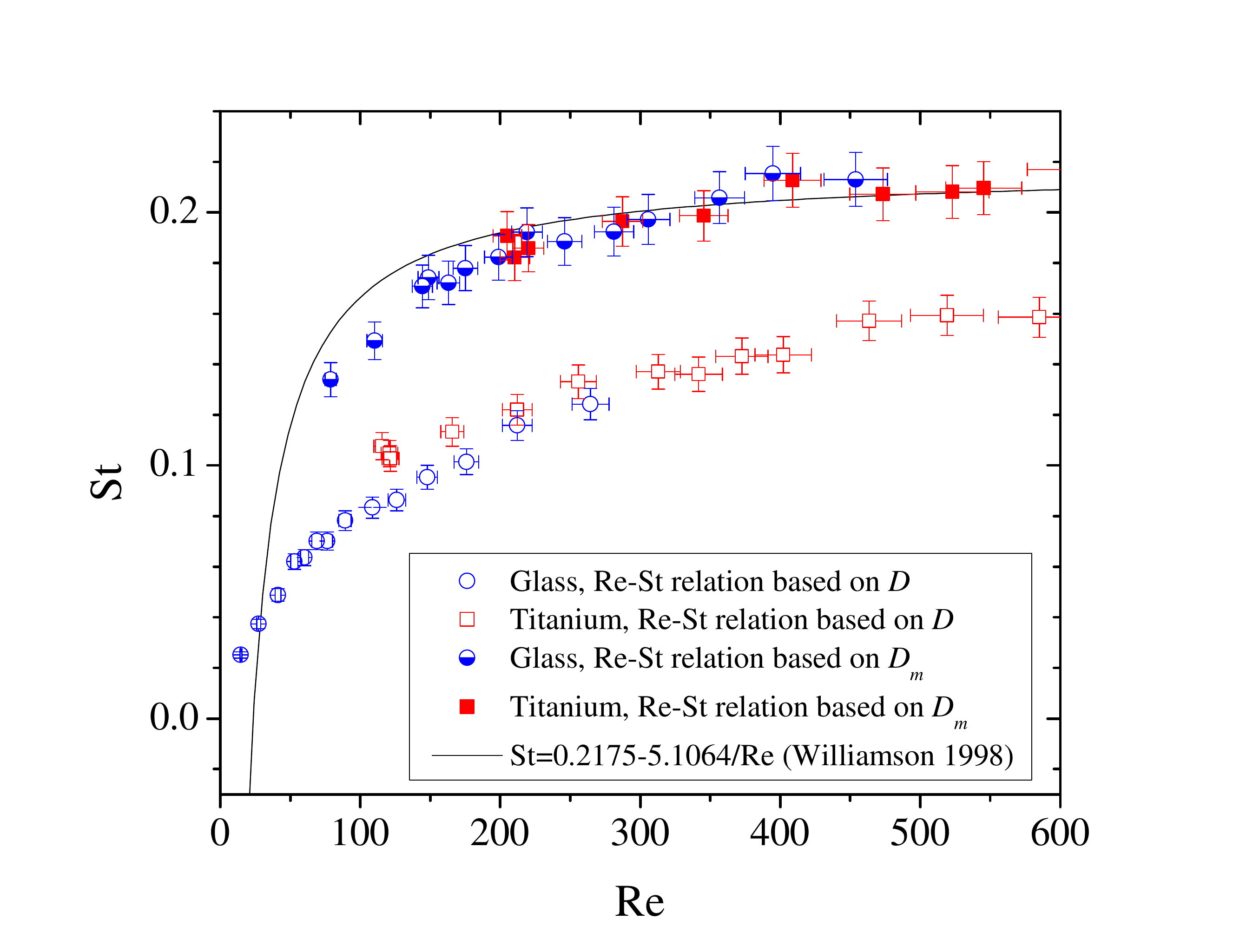}
\par
\end{centering}
\caption{
The $\Strouhal-\Reynolds$ relationships based on $D$ (open symbols) and $D_m$ (closed symbols).
There exists a slight mismatch between $\Strouhal$ acquired using glass and titanium around $\Reynolds=[100,200]$.
However, the mismatch reduces below the margin of error in $\Strouhal_m-\Reynolds_m$ relationship. 
The solid curve shows the $\Strouhal-\Reynolds$ relationship in three dimensions.
\label{fig:St-Re}}
\end{figure}

In Fig. \ref{fig:St-Re}, we plot the Strouhal-Rynolds number relationships as calculated based on $D$ and $D_m$.
Similar to our observation of the wavelength with respect to $D$ and $D_m$, the Strouhal number is not continuously measured with respect to the Reynolds number when these non-dimensional numbers are defined using $D$. 
Near $\Reynolds\sim[100,200]$, the measurements using glass and titanium deviate from each other outside the margin of error. 
As $\Reynolds$ increases, the two measurements overlaps within the margin of error.
However, if we define the non-dimensional numbers using $D_m$, the two measurements are statistically indistinguishable.
This observation assures our observation of the wavelength in that the meniscus impacts the formation of vorticity.

We note that a datum moves toward upper-right side if the non-dimensional numbers are defined using $D_m$.
Both $\Reynolds$ and $\Strouhal$ are proportional to $D$, and if $D$ is substituted by $D_m$, which is always larger than $D$, both numbers enlarge. 
When $D$ is small, i.e., when $\Reynolds$ and $\Strouhal$ are relatively small, this shift cam be substantial. 

Finally, we suspect that the two-dimensional $\Reynolds$-$\Strouhal$ relationship has been measured to be different from its three-dimensional counterpart partly because of the effect of meniscus. 
In many studies, the two-dimensional vortex streets are observed to behave differently from the three-dimensional vortex streets.
For example, different parameters of $\Reynolds$-$\Strouhal$ relationship were measured, and even the onset Reynolds number did not match.
However, as shown in Fig. \ref{fig:St-Re}, the $\Reynolds$-$\Strouhal$ relationship based on $D_m$ matches the Rayleigh's relation for laminar vortex streets in three-dimension, i.e., $\Strouhal=0.2175-5.1064/\Reynolds$ \cite{Ponta:2004we,Williamson:1988vf}.
This matching suggests that the difference between two- and three-dimensional laminar vortex streets is, at least partly, due to an experimental artifact.
It also suggests that some of small Reynolds number physics in soap film flows, such as the onset Reynolds number, needs further investigation.

\section{Summary}

To summarize, we investigated the effect of menisci to the downstream formation of vortex streets when an external object is inserted into the soap films.
In our experiment, we inserted objects composed of different materials, including aluminum, titanium, and glass, into the soap films and measured the size of the meniscus using a long-distance microscope.
The key findings of this study are two-fold.

First, we found that the meniscus is much smaller than the capillary length.
In our experiments, the size of the meniscus is initially proportional to the size of the object but reaches to asymptotic value.
The asymptotic value is approximately 0.2 to 0.3 mm, depending on the material. 
However, this measurements is much smaller than the capillary length of either water (2.7 mm) or soapy water (1.7 mm).
Even if we take account of the contact angles, the measurement is approximately five factor smaller smaller than the estimation of the dimensional analysis.

Second, we found that the meniscus acts as an added length scale to the size of object.
Data from different materials collapses into a single curve if we interpret the vortex street by adding the size of the meniscus to the size of object. 
This observation implies that menisci affect the downstream vortex structure, and this added length effect may need to be considered for the accurate assessment of data of future soap film studies.

However, such added length effect is most eminent only when the size of the object is comparable to the size of meniscus.
The meniscus does not exceed 0.2 mm for all cases tested.
Considering that most soap film studies use an object size $\sim\mathcal{O}(1)$ mm, any errors to be caused by the meniscus may be limited.

\bibliography{vortex2021}

\end{document}